\documentclass[11pt,twoside]{article}

\usepackage{asp2006}
\usepackage{epsf}
\usepackage{psfig}
\usepackage{lscape}

\begin{document}
\title{The Evolution of Primordial Circumstellar Disks}  
\author{Lucas A. Cieza\footnotemark}  
\affil{Institute for Astronomy, University of Hawaii at Manoa, HI, USA}    

\begin{abstract}
Circumstellar disks are an integral part of the star formation process and the sites 
where planets are formed. Understanding the physical processes that drive 
their evolution, as disks evolve from optically thick  to optically thin, is crucial 
for our understanding of planet formation.
Disks evolve through various processes including accretion onto the star, dust settling 
and coagulation, dynamical interactions with forming planets, and photo-evaporation. 
However, the  relative importance and timescales of these processes are 
still  poorly understood. In this review, I summarize current models of 
the different processes that control the evolution of primordial circumstellar 
disks around low-mass stars (mass $<$ 2 M$_{\odot}$). I also  discuss recent 
observational developments on circumstellar disk evolution with a focus on new 
\emph{Spitzer} results on transition objects.
\end{abstract}


\footnotetext{Spitzer Fellow}

\section{Introduction}\label{intro}  

It is currently believed that virtually all stars form surrounded by a circumstellar disk, 
even if this disk is sometimes very short lived. This conclusion follows from 
simple conservation of angular momentum arguments and is supported by mounting evidence.  
This evidence ranges from the excess emission, extending from the near-IR (Strom et al. 1989) to the 
sub-millimeter (Osterloh $\&$ Beckwith 1995),  that is observed in most young (age $<$ 1 Myr) 
pre-main-sequence (PMS) stars  to direct 
\emph{Hubble} images of disks (McCaughrean \& O'Dell 1996) seen as silhouettes in front of the 
Orion Nebula. Also, even though there is not \emph{direct} evidence 
that planets actually grow from circumstellar 
material, it has become increasingly clear that they are in fact the birthplaces of planets since 
their masses, sizes, and compositions are consistent with the theoretical
minimum-mass solar nebula (Hayashi 1981). Recently, the discovery of exo-planets orbiting 
nearby main sequence stars has confirmed that the formation of planets is a common process and 
not a rare phenomenon exclusive to our Solar System. Thus, any theory of planet formation should 
be robust enough to account for the high incidence  of planets and cannot rely on special 
conditions or on unlikely processes to convert circumstellar dust and gas into planets.  

        Star and planet formation are intimately related. Standard low-mass star formation 
models (e.g. Shu et al. 1987) describe the free fall collapse of a slowly rotating molecular 
cloud core followed by the development of a hydrostatic proto-star surrounded by an envelope 
and a disk of material supported by its residual angular momentum. 
This early phase is expected to occur on a timescale of about $10^5$ years 
(Beckwith 1999) and results in an optically revealed classical T Tauri star (CTTS, low-mass  
PMS star that shows clear evidence for accretion of circumstellar material). 
This stage is characterized by intense accretion onto the star, strong winds, and 
bipolar outflows. As the system evolves, presumably into a weak-lined T Tauri star 
(WTTS, PMS star mostly coeval with CTTSs but that do not show evidence for accretion), 
accretion ends, and the dust settles into the mid-plane of the disk where the solid particles 
are believed to stick together and to grow into planetesimals as they collide. 
Once the objects reach the kilometer scale, 
gravity increases the collision cross-section of the most massive planetesimals, and runaway 
accretion occurs (Lissauer 1993). In the standard core accretion model (e.g., Pollack et al. 1996), 
massive enough proto-planets still embedded in the disk can accrete the remaining gas and 
become giant planets. The early stages are the most uncertain, and many people suspect that 
grains will not grow into planetesimals by collisions alone at the rate necessary to go 
through all stages of giant planet formation before the gas nebula has dissipated.
Also, even assuming that grains can grow into macroscopic bodies at the
necessary rate, it is still doubtful whether they can form large planetesimals. 
Once objects reach the meter-size scale, they are expected to rapidly spiral inward
due to the dynamical interactions with the gas in the disk, which rotates at a slightly 
sub-Keplerian velocity because it is supported  against gravity by pressure in addition
to the centrifugal force.

However, since current statistics of extra-solar planets  
indicate that giant planets are common, the difficulties of the standard core accretion 
model have led some researchers (e.g., Boss 2000) to revisit an  alternative 
planet formation mechanism that had been put aside for several decades, namely, gravitational 
instability. In the gravitational instability scenario,  giant planets form through the direct 
gravitational collapse of a massive unstable disk over timescales $<$ 10$^{4}$ yrs. In some
hybrid models (e.g.,  Youdin $\&$ Shu 2002), solid particles settle to the mid-plane of the 
circumstellar disk and form a dense gravitationally unstable sub-disk. In this manner, 
planetesimals form from the gravitational collapse of the material in the mid-plane.  The collision 
of planetesimals leads to runaway accretion, and the process continues in a way  analogous to the 
core accretion model.

Thus, while the existence of planets around a significant fraction of all the stars is 
considered verified, the precise mechanisms through which planets form still remain largely 
unknown. So far, none of the proposed theories have proven satisfactory. On one hand, the 
standard model of continuous accretion of solid particles relies on doubtful sticking properties 
of rocks and on unknown processes to prevent the migration of meter-size objects. 
On the other hand, gravitational instability relies on unproven mechanisms to enhance the 
surface density of the disk's mid-plane to trigger the process of  planet formation. 
Clearly, more observational constraints are necessary to help the theoretical work 
on planet formation to proceed forward. 

\begin{figure}[!ht]
\plotone{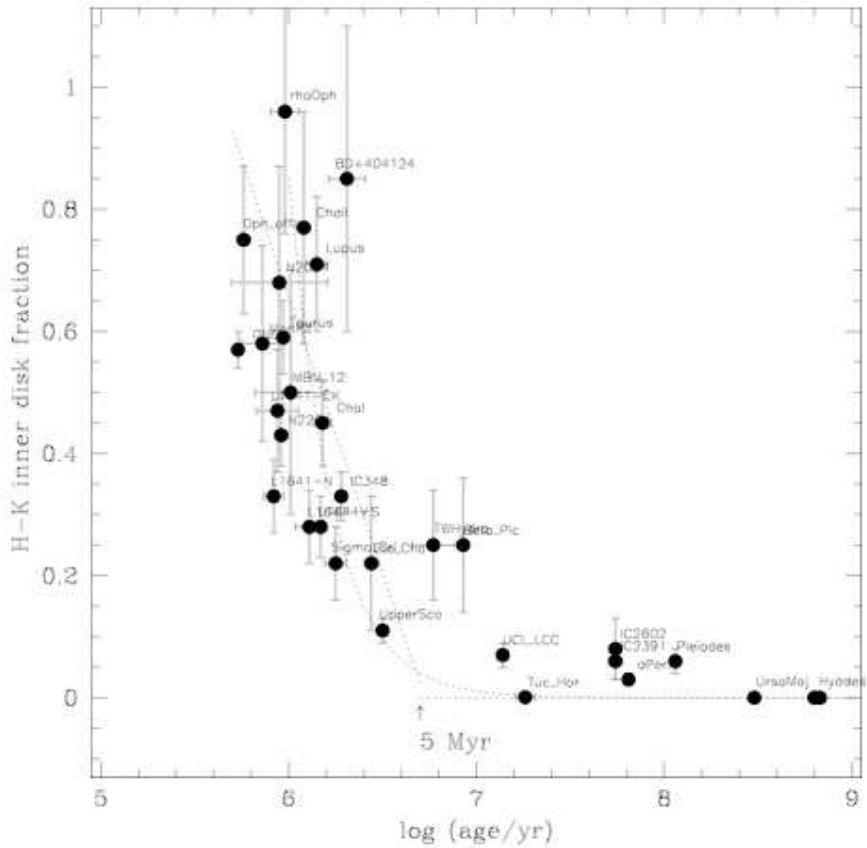}
\caption{
Inner accretion disk fraction vs. stellar age inferred from H-K excess measurements,
binned by cluster or association, for $\sim$3500 stars from the literature. 
(Figure taken from Hillenbrand 2006).
}
\end{figure}

\section{Dissipation Timescales}\label{timescales}

Until the Atacama Large Millimeter Array (ALMA) becomes operational in the next decade, 
direct detection of forming planets will remain beyond our capabilities, but circumstellar 
disks are easier to detect and study because the  surface area of a planetary mass dispersed into 
small grains is \emph{many} orders of magnitude greater than the surface area of a planet. 
For this reason, much of the information about the properties of disks such 
as size, mass, density, and evolution timescales has been obtained by observing the 
thermal emission of the dust gains. These particles absorb and re-radiate the light 
mostly in the 1 $\mu$m - 1 mm range. Since the temperature of the disk decreases with 
the distance from the central star, different wavelengths probe different disk radii.

One of the most important quantities on disk evolution is the lifetime
of the disk itself, not only because it establishes the relevant timescale of the
physical processes controling disk evolution, but also because it sets a limit for 
the time available for planet formation. In what follows, I review the constraints on 
the dissipation timescale of different regions of the disk obtained from different surveys, 
performed at wavelengths ranging from the near-IR to the sub-millimeter.

\subsection{Inner Disk}   

Since there is an almost 1 to 1 correlation between the presence of near-IR excess (1-5 $\mu$m) 
and the occurrence of spectroscopic signatures of accretion (Hartigan et al. 1995), it is possible to 
investigate the lifetime of inner accretion disks (r $<$ 0.05-0.1 AU) by studying the 
fraction of stars with near-IR excess as a function of stellar age. Early studies of nearby 
star-forming regions (e.g., Strom et al. 1989) found that 60-80$\%$ of the stars younger than 
1 Myr present  measurable near-IR excesses, and that just 0-10$\%$ of the stars older than 10 Myr do so.
It has been argued that individual star-forming regions lack the intrinsic age spread necessary 
to investigate disk lifetimes from individually derived ages 
(Hartmann 2001). However, similar disk studies,  based on the disk frequency in clusters 
with different mean ages and extending to the 3.4 $\mu$m L-band (Haisch et al. 2001, 
Hillenbrand 2006), have led to results similar to those presented by previous groups. 
It is now well established that the frequency of inner accretion disks steadily decreases
from $<$1 to 10 Myr. This decrease is illustrated in Figure 1 for a sample of
over 3500 PMS stars in nearby clusters and associations.  The disk fractions 
are consistent with  \emph{mean} disk lifetimes on the order of 2-3 Myr and a wide 
dispersion: some objects lose their inner disk at a very early age, even before they become 
optically revealed and can be placed in the HR diagram, while other objects retain their 
accretion disks for up to 10 Myr.

\subsection{Planet-forming Regions of the Disk}

Disk lifetime studies based on near-IR excesses always left room for the possibility 
that stars without near-IR excess had enough material to form planets at larger radii 
not probed by near-IR  wavelengths. The IRAS and ISO observatories had the appropriate 
wavelength range to probe the planet-forming regions of the disk (r $\sim$0.05-20 AU) 
but lacked the sensitivity needed to detect all but the strongest  mid- and far-IR excesses 
in low-mass stars at 
the distances of nearest star-forming regions. \emph{Spitzer} provides, for the first time, 
the wavelength coverage and the sensitivity needed to detect very small amounts of dust in the 
planet-forming regions of a statistically significant number of low-mass PMS stars.
Recent \emph{Spitzer} results (e.g., Padgett et al. 2006; Silverstone et al. 2006; Cieza el al. 2007) 
have shown that PMS stars lacking near-IR excess are also very likely to show 24 $\mu$m
fluxes consistent with bare stellar photospheres and that optically thick primordial
disks are virtually non-existent beyond an age of 10 Myr. Cieza et al. (2007)
also find that over 50$\%$ of the WTTS younger than $\sim$1-2 Myr  show no evidence 
for a disk, suggesting that the inner $\sim$10 AU of a significant fraction of  
all PMS stars becomes extremely depleted of dust (mass $<$ $10^{-4}$ M$_{\oplus}$)
by that early age. 
 
\subsection{Outer Disk}

Recent sub-millimeter results extend the conclusions on the survival time
of the material in the inner disk (r $<$ 0.1 AU) and the planet-forming region 
of the disk (r $\sim$ 0.5--20 AU) to the outer disk (r $\sim$50-100 AU). 
Andrews $\&$ Williams (2005, 2007) study over 170 Young Stellar Objects (YSOs) 
in the Taurus and Ophiuchus molecular clouds and find that $<$ 10$\%$ of 
the objects lacking inner disk signatures are detected at sub-mm wavelengths. 
Given the mass  sensitivity of their survey (M$_{DISK}$ $\sim$
$10$$^{-4}$--10$^{-3}$ 
M$_{\odot}$), they conclude that the dust in the inner and the outer disk dissipates 
nearly simultaneously.

\subsection{Implications of Disk Lifetimes}\label{lifetimes} 

Taken together, the results of the surveys from the near-IR to the sub-millimeter imply 
that a sizable fraction of all PMS stars lose their disks completely before the star reaches 
an age of $\sim$1 Myr, while some CTTSs maintain a healthy accretion disk for up to $\sim$10 Myr,
which seems to be the upper limit for the survivability  of primordial disks. 
The reason why, within the same molecular cloud or stellar cluster, some primordial 
disks survive over 10 times longer than others is still unknown. The spread in disk lifetimes 
is likely to be related to the wide range of initial conditions,  the planet formation process, 
and/or the effect of unseen companions. 


Since recent core accretion models (e.g., Alibert et al. 2004) can accommodate planet 
formation within 
10 Myr, the disk survivability limit alone can not distinguish between the competing planet 
formation mechanisms, core accretion and gravitational instability. On one hand, it is 
possible that planets form through core accretion around the few CTTS disks that manage 
to survive for $\sim$5-10 Mys. On the other hand, it is also possible that the youngest 
WTTSs \emph{without} a disk are objects that have already formed  planets through 
gravitational instability.  Thus, establishing the incidence of planets around ``young WTTSs'' 
and ``old CTTSs'' could provide a crucial observational discriminant between the core 
accretion and the gravitational instability models. This goal is is one 
of the central objectives of the ``Young Stars and Planets''  Key Project of the Space 
Interferometry Mission (Beichman et al. 2002), a mission for which launch has
unfortunately been deferred indefinitely at the time of writing.

\begin{figure}[!ht]
\plotone{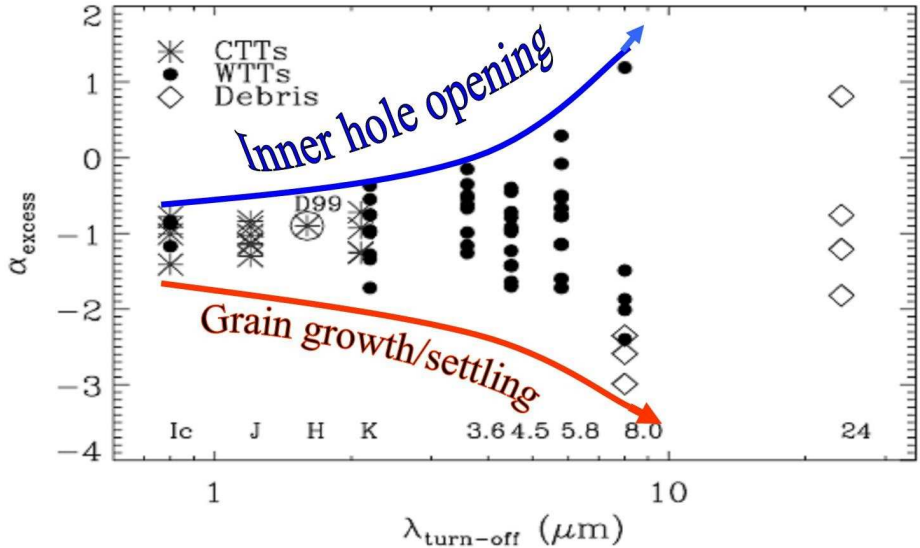}
\caption{
Distribution of excess slopes, $\alpha$$_{excess}$, vs. the wavelength at which the 
infrared excess begins, ($\lambda$$_{turn-off}$, for the sample of WTTSs (filled circles) 
from Cieza et al. (2007), 
a sample of CTTSs in Chamaeleon (asterisk) from Cieza et al. (2005), 
the median SED of CTTSs in Taurus (marked as D99) from D'Alessio et al. (1999), 
and a sample of debris disks (diamonds) from Chen et al. (2005). 
The diagram shows a much larger spread in 
inner disk properties of WTTSs with respect to those of CTTSs.
The spread in disk properties suggests that different physical 
processes dominate on different objects. Objects at the top of the figure
are consistent with disk evolution dominated by the opening of an inner hole,
while objects at the bottom of the figure are more consistent with 
evolution dominated by dust seetling. 
 (Figure adapted from Cieza et al. 2007). }
\end{figure}

\section{Transition Timescales and Transition Objects}\label{trans}

The fact that very few objects lacking near-IR excess show mid-IR or sub-millimeter excess emission  
implies that, once  accretion stops, the entire disk dissipates very rapidly. Based on the relative 
numbers of these objects, the transition/dissipation timescale is estimated to be  $<$ 0.5 Myr 
(Skrutskie et al. 1990; Wolk $\&$ Walter 1996; Cieza et al. 2007). From the observational point 
of view, this short transition timescale 
means that the vast majority of PMS stars in any given population are either accreting CTTSs with excess 
emission extending all the way from the near-IR to the sub-millimeter or bare stellar photospheres.
This also means that any T Tauri star whose SED does not look like a typical CTTS or a bare 
stellar phothosphere can be broadly characterized as a ``transition object''. It should be noted, 
however, that precise definitions of what constitutes a transition object found in the disk 
evolution literature are far from homogeneous.  

Since there is a very strong correlation between accretion and the presence of near-IR 
excess (Hartigan et al. 1995), WTTS very rarely show near-IR excess; therefore, any 
WTTS that do show IR excesses at longer wavelengths is a transition disk 
according to the broad definition stated above.  Padgett et al. (2006) and Cieza et al. (2007) show 
that the few WTTS that do have a disk present a wide diversity of SED morphologies and disk to 
stellar luminosity ratios that bridge the gap observed between the CTTS and the debris disk regime.  
The ratio of the disk luminosity to the stellar luminosity, $L_{DISK}/L_{*}$, is a measurement of 
the fraction of the star's radiation that is intercepted and re-emitted by the disk plus any accretion 
luminosity. This quantity is intimately related to the evolutionary status of a circumstellar disk.
On the one hand, the primordial, gas rich, disks around CTTSs have typical $L_{DISK}/L_{*}$ values 
$>10-20$$\%$, mostly because they have optically thick disks that intercept $\sim$10-20$\%$ of the stellar 
radiation. On the other hand, gas poor debris disks have optically thin disks that intercept a much 
smaller fraction of the star's light and thus have $L_{DISK}$/$L_{*}$ values that  range from $10^{-3}$ to 
$10^{-6}$ (Beichman et al. 2005).

The fact that WTTSs have $L_{DISK}/L_{*}$ values intermediate between those of CTTSs
and debris disks suggests that they are an evolutionary link between these two well studied stages.
These rare WTTS disks are key for our understanding of disk evolution because they seem to 
trace the dissipation process as it rapidly occurs across the entire disk. 
The wide range of SED morphologies seen in WTTS disks suggests that the disk dissipation process 
does not follow the same path for every star. 
The diversity of SED morphologies of WTTS disks is quantified in Figure 2, which shows the slope of 
the IR excess, $\alpha$$_{excess}$,
against the wavelength at which the IR becomes signicant, $\lambda$$_{turn-off}$.
Objects in the top right of the figure are WTTS that have lost their short wavelength excess while 
keeping strong excesses at longer wavelengths. The SEDs of these objects are consistent with 
the formation of an inner hole in their disks (Calvet et al. 2002, 2005). This possible 
``inner hole opening'' path is illustrated by the SEDs of the objects shown in the top row 
of Figure 3. 

Objects in the bottom right of Figure 2  are WTTS in which the IR excess seems to have decreased 
simultaneously at every wavelength. These objects are more consistent with an evolution 
dominated by grain growth and dust settling. As the grains grow and settle into the mid-plane 
of the disk, the flared disk becomes flatter and hence intercepts a smaller fraction of the 
stellar radiation.  As a result, a smaller excess is expected at every mid-IR  wavelength 
(Dullemond $\&$ Dominik 2004). This possible ``dust settling'' path is illustrated by the SEDs 
shown in the bottom row of Figure 3. 

\begin{figure}[!ht]
\plotone{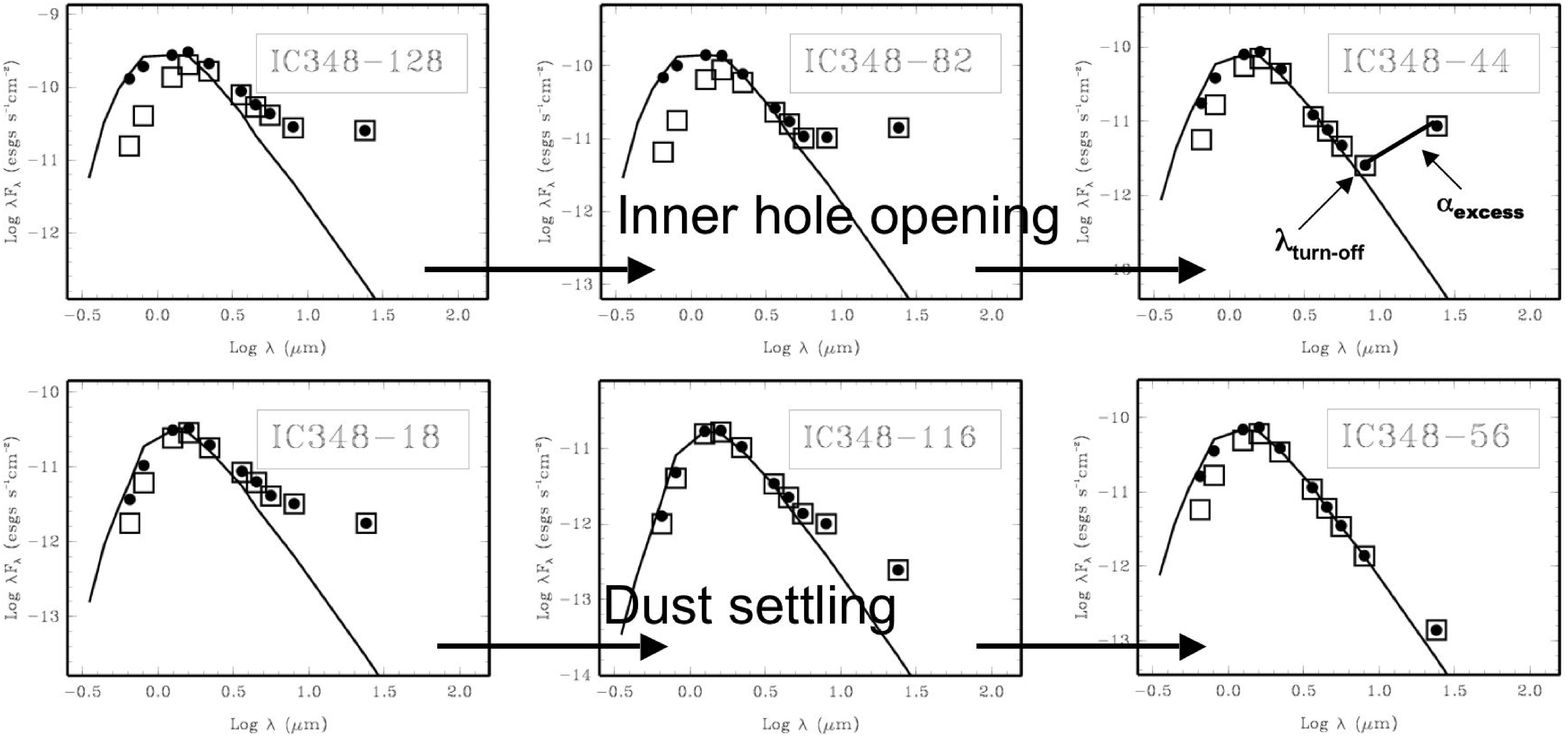}
\caption{
The SEDs of several WTTSs from Figure 2. 
The open boxes are observed fluxes, the  filled
circles are extinction corrected values,  and the 
solid lines represent the stellar photospheres.
The SEDs illustrate two possible evolutionary
sequences, one dominated by the opening 
of an inner hole (top row) and the other
dominated by dust settling (bottom row).
}
\end{figure}

\subsection{The Full Diversity of Transition Disks}

It is important to note that while most CTTSs occupy a very restricted region of
the $\alpha$$_{excess}$ vs. $\lambda$$_{turn-off}$ diagram, some CTTSs do show 
transition SEDs similar to those of WTTS disks. The best known examples of such CTTSs 
include TW Hydra (Calvet et al. 2002), DM Tau, and GM Aur (Calvet et al. 2005). 
These systems likely represent yet another path for disk dissipation, a path in 
which the inner regions of the disk become depleted of dust while gas is still accreting 
onto the star. 

The standard YSO evolutionary class proposed by Lada (1987) and extended by 
Greene et al. (1994) is based on the slope, $\alpha$, of the SED between 2 and 25 $\mu$m.
However, this classification scheme can only capture a single evolutionary path in 
which $\alpha$ decreases monotonically as young stellar objects evolve from Class I into Class 
II, and then into Class III objects. The mid-IR data points from \emph{Spitzer}'s camera IRAC 
(3.6, 4.5, 5.8, and 8.0 $\mu$m) 
sample the SEDs of YSOs at intermediate wavelengths between 2 and 25 $\micron$ and reveal 
that transition objects present a much larger diversity of SED morphologies than what can be 
described by the $\alpha$ classification. The $\alpha$$_{excess}$--$\lambda$$_{turn-off}$ 
classification seems more appropriate when abundant mid-IR broad-band photometry is available, 
but it still does not uncover the full range of SED shapes presented by transition 
objects.  

\emph{Spitzer}'s Infrared Spectrograph (IRS), which allows us to sample the SED of YSOs 
at hundreds of wavelengths between 5 and 38 $\mu$m, has recently revealed a new family
of transitions objects whose SEDs show a distinctive ``dip'' around $\sim$10-20 $\mu$m (Brown et al. 2007). 
The SEDs of these objects can be modeled as disks having wide gaps with small inner radii (0.2-0.8 AU) 
and large outer radii (15-50 AU).  The presence of small IR excess shortward of 10 $\mu$m in 
these objects requires the presence of small amounts of dust (M $\sim$ 10$^{-6}$$M_{\odot}$) 
in the inner disk, unlike objects such as Coku Tau/4 (D'Alessio et al. 2005) and DM Tau 
(Calvet et al. 2005), whose inner holes seem to be completely depleted of dust. 
Since the family of disks with SEDs suggestive of gaps contains both accreting objects 
(e.g., LkH$\alpha$ 330 and HD 135344) and not accreting objects (SR 21 and T Cha), 
it significantly expands the diversity of transition disks.

\section{Disk Evolution Processes}

The processes that drive disk evolution and the different
pathways that disks follow as they evolve from optically thick primordial disks 
to optically thin debris disks are critical for our understanding of planet 
formation. These processes include accretion onto the star, photo-evaporation, 
dust settling and coagulation, and dynamical interactions with forming 
planets. In what follows, I review the main processes believed to control 
the evolution of circumstellar disks around low-mass stars. 

\subsection{Viscous Evolution}

To first order, the evolution of primordial disks is driven by viscous 
accretion. Circumstellar material can only be accreted onto the star if 
it loses angular momentum.  Conservation of angular momentum implies
that, while most of the mass in the disk moves inward, some material 
should move outward, increasing the size of the initial disk. 
The source of the viscosity required for disk accretion and  
the mechanism by which angular momentum is transported remain 
a matter of intense debate. As a result, most viscous evolution 
models describe viscosity, $\nu$, using the $\alpha$ parameterization
introduced by Shakura $\&$ Sunyaev (1973), according to which 
$\nu$ = $\alpha$$H_p$$C_s$, where $H_p$ is the pressure scale
height of the disk and $C_s$ is the isothermal sound speed. 
The parameter $\alpha$ hides the uncertainties associated with the 
source of the viscosity and is often estimated to be of the order of 0.01.

Viscous evolution models (Hartmann et al. 1998; Hueso $\&$ Guillot 2005) 
are broadly consistent with the observational constraints for disk 
masses, disk sizes, and accretion rates as a function of time;  however,
they also predict a smooth, power-law, evolution of the disk properties.
This smooth disk evolution is inconsistent with the very rapid 
disk dissipation ($\tau$ $<$ 0.5 Myr) that usually occurs after 
a much longer disk lifetime. 
Pure viscous evolution models also fail to explain the variety 
of SEDs observed in transition objects dicussed above.
These important limitations of the viscous evolution models
suggest that they are in fact just a first-order approximation of 
a much more complex process. 

\subsection{Photo-evaporation}
 
Recent disk evolution models, known as ``UV-switch'' models, combine viscous evolution 
with photo-evaporation by the central star (Clarke et al. 2001; 
Alexander et al. 2006) and are able to account for both the disk lifetimes 
of several million years and the short disk dissipation timescales  ($\tau$ $<$ 0.5 Myr). 
According to these models, extreme ultraviolet (EUV) photons originating in the accretion 
shock close to the stellar surface ionize and heat the circumstellar hydrogen to 
$\sim$10$^4$ K. Beyond some critical radius, the thermal energy of the 
ionized hydrogen exceeds its escape velocity and the material 
is lost in the form of a wind. 

At early stages in the evolution of the disk, the accretion 
rate dominates over the evaporating rate and the disk undergoes 
standard viscous evolution: material from the inner disk is accreted 
onto the star, while the outer disk behaves as a reservoir that
resupplies the inner disk, spreading as angular momentum is transported
outwards. Later on, as the accretion 
rate drops to the photo-evaporation rate, 
the outer disk is no longer able to resupply the inner disk with 
material. At this point, the inner disk drains on a viscous timescale and
an inner hole is formed in the disk. Once this inner hole 
has formed, the EUV radiation very efficiently photo-evaporates 
the inner edge of the disk and the disk rapidly dissipates from 
the inside out.  Thus, the UV-switch model naturally accounts for 
the lifetimes and dissipation timescales of disks as well as for 
SEDs of some PMS stars suggesting the presence of large inner holes.

Photo-evaporation, however, is not the only mechanism that has been
proposed to explain the large  opacity holes of some 
circumstellar disks. Dynamical interactions with planets and 
even grain growth can deplete the inner disk of small grains and 
result in SEDs virtually indistinguishable from those expected 
for photo-evaporating disks.

\begin{figure}[!ht]
\plotone{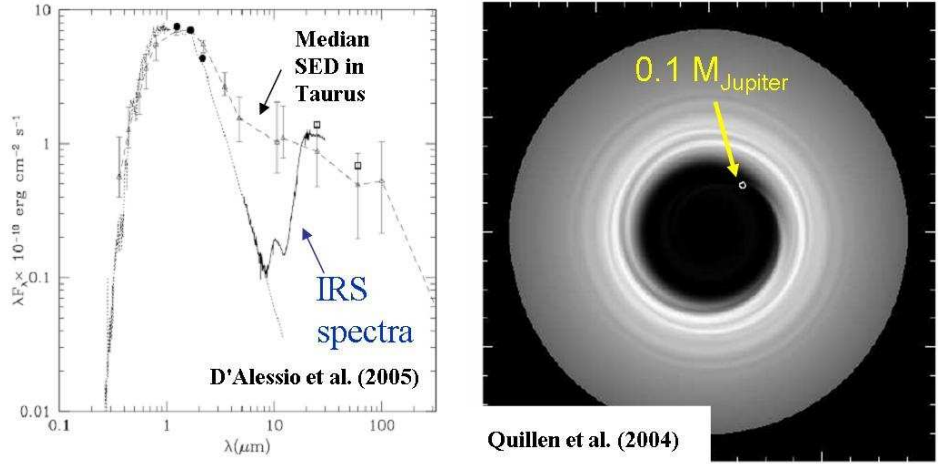}
\caption{The left panel shows the spectral energy distribution of CoKu Tau/4 (the 
IRS spectra is from Forrest et al. 2004), compared to the median SED of CTTSs 
in Taurus (triangles and dashed line). The error bars of the median 
points represent the upper and lower quartiles of the Taurus CTTS SEDs 
(D'Alessio et al. 1999). (Figure adapted from
D'Alessio et al. 2005). The right panel shows an hydrodynamic simulation of 
a planet disk system with a planet mass of 0.1M$_{Jupiter}$. 
This putative planet is suggested to be responsible for the evacuation of 
the inner disk of Coku Tau/4. (Figure taken from Quillen et al. 2004).
}
\end{figure}

\subsection{Dynamic Interaction with Planets}

Since theoretical models of the dynamical interactions of 
forming planets with the disk (Lin $\&$ Papaloizous 1979, 
Artymowicz $\&$ Lubow 1994) predict the formation of inner holes 
and gaps, planet formation quickly became one of the 
most exciting explanations proposed for the inner holes of 
transition disks (Calvet et al. 2002; D'Alessio et al. 2005). 
Using a combination of hydrodynamical simulations and  
Monte Carlo radiative trasport, Rice et al. (2003) model the 
broad-band SED of the transition disk around the CTTS GM Aur. 
They argue that the $\sim$4 AU inner hole they infer for the 
disk could be produced by a $\sim$2 M$_{Jupiter}$ mass planet 
orbiting at 2.5 AU. Modeling of the \emph{Spitzer}-IRS data 
of GM Aur places the inner wall of the hole 
significantly farther away from the star, at 24 AU (Calvet et al. 2005);
nevertheless, the planetary origin of the gap still remains one of 
the favored hypotheses (Najita et al. 2007). 

Similarly, hydrodynamic simulations (Quillen et al. 2005)
suggest that the 10 AU hole inferred for the disk around
the WTTS Coku Tau/4 (Figure 4, left panel) could be produced 
by the presence of a $\sim$0.1 M$_{Jupiter}$ mass planet 
orbiting very close to the edge of the hole's wall (Figure 4, right panel).  
However, the very low accretion rate and the low disk mass render
the disk around Coku Tau/4 one of the prime candidates for 
ongoing photo-evaporation (Najita et al. 2007).

Given the extreme youth of some of the transition objects that  
have disks with inner holes (age $\sim$ 1 Myr), the confirmation 
of the planetary origin of these holes would place very 
strong constraints on the time required for the formation of 
giant planets.
In the not too distant future, ALMA will provide the 
sensitivity and angular resolution necessary to image
the putative planets responsible for the observed holes
(Narayanan et al. 2006).     

\subsection{Grain Growth}

Even though the formation of a giant planet can 
account for the inner holes of some transitional disks, the planet 
formation process is not required to be far along in order to produce a  
similar effect in the SED of a circumstellar disk. In fact, once 
primordial sub-micron dust grains grow into somewhat larger bodies 
(r $\gg$ $\lambda$), most of the solid mass never interacts with 
the radiation, and the opacity function, $k_{\nu}$ ($cm^2$/gr), 
decreases dramatically. Dullemond $\&$ Dominik (2005) model
the settling and coagulation of dust in disks and investigate
their effect on the resulting SEDs. They find that grain growth 
is a strong function of radius, it is more efficient in the inner
regions where the surface density is higher and the dynamical 
timescales are shorter, and hence can produce opacity holes. 
As a result, grain growth, the very first stage of planet formation,
can in principle mimic the effect of a fully grown giant planet
on the SED of a YSO.
Dullemond $\&$ Dominik (2005) also find that, in their models,
grain growth is \emph{too efficient} to be consistent with the
observed persistence of IR excess over a timescale of a few Myrs. 
They conclude that small grains need to be replenished by fragmentation 
and that the size distribution of solid particles in a disk is likely to
be the result of a complicated interplay between coagulation  
and fragmentation.  As such, the overall importance of the 
dust coagulation process on the evolution of SEDs is still 
not well understood.  

\subsection{Inside-out Evacuation by Magnetorotational Instability}

The formation of an opacity hole, due to grain growth or the presence of 
a planet, can have not only a dramatic effect on the appearance of the SED of
a transition object, but also on disk evolution itself.  
Chiang $\&$ Murray (2007) recently propose that, once 
an opacity hole of the order of 1-10 AU is formed in a disk, the inner 
rim of the disk will be rapidly drained from the inside out
due to the onset of the magneto rotational instability (MRI).
The MRI is one of the mechanisms that has been historically proposed 
to explain the accretion process in circumstellar disks.  
Nevertheless, it is believed that the gas in CTTSs is usually 
too cold and too weakly ionized for the MRI to be efficient 
(Hartman et al. 2006). Typical CTTS disks are too dusty for the 
stellar X-rays to penetrate deep enough into the disk to
ionize material to the level required by the MRI. 
Chiang $\&$ Murray (2007) argue that an opacity hole in a 
disk produces the right conditions for the MRI to 
be activated and sustained. According to their models, the
wider the rim, the larger the mass of the MRI-active 
region of the disk and the higher the accretion rate.  
Therefore, once the MRI is activated, the entire inner disk
is rapidly evacuated, while the outer disk is 
photo-evaporated by UV radiation. 

\subsection{Models vs Observations}

To date, most disk evolution models have treated each one of the processes
discussed above independently of each other. In reality, it is clear that  all these processes 
are likely to operate simultaneously and interact with one another. Hence, 
the relative importance of these processes for the overall evolution of disks 
still remains to be established. The diversity of SEDs morphologies discussed 
in Section 3 strongly suggests that disk evolution can follow 
different paths, each one of which is likely to be dominated by one or more 
different physical processes.

In order to investigate the relative incidence of different evolutionary 
paths, Najita et al. (2007) compare the observed accretion rates and disks masses of 
transition objects, which they define as objects with weak or no excess shortward of 
10 $\mu$m and strong excess at longer wavelengths, against the values expected 
for several disk evolution models. They find that transition disks occupy a 
restricted region of the accretion rate vs. disk mass plane. In particular, they 
find that transition objects tend to have significantly lower accretion rates, 
for a given disk mass, than non-transition objects and also tend to
have larger median disk masses than regular T Tauri disks. 

Najita et al. (2007) argue that most of the objects in their sample have properties consistent
with a scenario in which a jovian mass planet has created a gap that isolates the 
inner disk from a still massive outer disk (Lubow $\&$ D'Angelo 2006; 
Varniere et al. 2006) and suppresses accretion onto the star. 
They also argue that a minority of their objects have both small accretion rates 
and small disk masses that make them more consistent with the photo-evaporation
model. Finally, they propose that the lack of transition objects with large 
accretion rates and large disk masses implies that grain growth and the formation
of planetesimals in the inner disk is an unlikely explanation for the opacity
holes of transition disks. 
While very provocative, the results of Najita et al. (2007) are based on sample of 
only 12 transition objects. Fortunately, \emph{Spitzer} is likely to increase 
the number of known transition disks by an order of magnitude by the end of its 
mission.  Follow-up observations of these objects, required to obtain accretion
rates and disk masses, will soon allow to extend the study of Najita et al. (2007) 
to a much larger sample. 

\begin{figure}[!ht]
\plotone{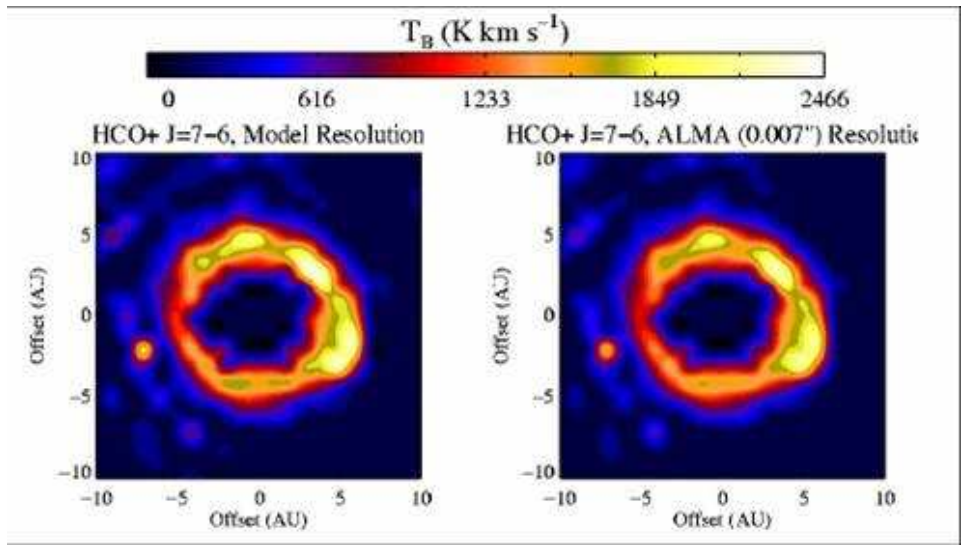}
\caption{
The left panel shows the synthesized image of the HCO+ (J=7-6) transition for a
hydrodynamic snapshot of the vicinity of a  self gravitating 
proto giant planet with a mass of 1.4$M_{Jupiter}$.
The right panel is a simulated ALMA image at 0.007 $''$ 
resolution (corresponding to the most extended baseline of ALMA, Bastian 2002) 
for a source at a distance of 140 pc.  The forming giant planet is at 8 o'clock 
in each panel. The intensity is in units of K km/s and is on a fixed scale 
for the entire figure. (Figure taken fron Narayanan et al. 2006).
}
\end{figure}

\section{Conclusions and the Future}

The many transition disks discovered by \emph{Spitzer} are likely to be 
the subjects of many follow-up observations and studies for years to come.
These are objects that are undergoing rapid transformations and 
have properties that place them  somewhere in between two much better
defined evolutionary states: those of regular CTTSs and debris disks. 
Understanding the physical processes responsible for the diversity of 
transition disks and its implications for planet formation will be 
one of main challenges for the field.

Another key outstanding disk evolution issue is the question of when the transition 
from the primordial to the debris disk stage actually occurs. As mentioned in Section 3, 
some of the WTTS disks have very small fractional luminosities and thus seem to be optically thin. 
These objects  \emph{could} be younger analogs of the $\beta$ Pic and AU Mic debris disks, and 
thus some of the youngest
debris disks ever observed. However, this interpretation depends on the assumption that these
young WTTS disks are gas poor. It is also possible that some of the WTTS disks have very low
fractional disk luminosities as a result of most of their grains growing to
sizes $\gg$10-20 $\mu$m, in which case they would still be primordial disks.
Since a real debris disks requires the presence of second generation of dust produced
by the collision of much larger objects, the ages of the youngest debris disks can
constrain the time it takes for a disk to form planetesimals.

It has even been proposed that detectable second generation dust is not produced
until Pluto-sized objects form and trigger a collision cascade (Dominik $\&$ Decin, 2003).
In that case, the presence of debris disks around $\sim$1-3 Myrs old stars
would have even stronger implications for planet formation theories.
The fact that, as discussed in Section 3, the incidence of 24 $\mu$m excesses 
around low-mass PMS stars with ages $\sim$10 Myr is $\sim$0 $\%$, is consistent with the idea
that a quiescent period exists between the dissipation/coagulation of the primordial dust
and the onset of the debris phenomenon. However, the confirmation of such a scenario will
require far-IR observations sensitive enough to detect, at the distance of the nearest
star-forming regions (150-200 pc),  debris disks as faint as those observed in the solar
neighborhood (10-30 pc). In the near future, \emph{Herschel} will provide such sensitivity.
It will also provide  the information on the gas content and grain size distribution of
optically thin disks around WTTS required to establish whether they represent the end of
the primordial disk phase or the beginning of the debris disk stage.

Even though much still remains to be learned about planet formation 
from the study of circumstellar disks, ultimately, we would like to be 
able to directly observe planets as they form. ALMA could 
achieve this milestone in the next decade. With its most
extended baseline, ALMA will have a resolution of 
0.007$''$ at 625 GHz. This corresponds to an angular resolution of less 
than 1 AU at the distance of nearby  star forming regions like Ophiuchus 
and Taurus.  Narayanan et al. (2006) model the molecular 
emission line from a gravitationally unstable protoplanetary
disk (Figure 5, left panel) and conclude that forming giant planets 
could be directly imageable using dense gas tracers such as  HCO+ (Figure 5, 
right panel). Such observations could provide the ultimate tests needed 
to distinguish among the competing theories of planet formation.

\acknowledgements 
Support for this work was provided by NASA through the \emph{Spitzer} 
Fellowship Program Under an award from Caltech.
I would like to thank Trent Dupuy, Jonathan Swift, and 
Michael Liu for their helpful suggestions.   


\end{document}